\documentstyle[preprint,prl,aps,epsfig]{revtex} 

\newcommand{\Tr}{{\rm Tr}\,}

\newcommand{\CN}{{\cal N}}

\newcommand{\CP}{{\cal P}}

\newcommand{\CZ}{{\cal Z}}
\newcommand{\CO}{{\cal O}}

\begin{document}
\draft
\title{\begin{flushright}
{\small\hfill AEI-099\\
\hfill hep-th/9902113}\\
\end{flushright}
Eigenvalue Distributions in Yang-Mills Integrals}
\author{Werner Krauth \footnote{krauth@physique.ens.fr }}
\address{CNRS-Laboratoire de Physique Statistique,
Ecole Normale Sup\'{e}rieure\\
24, rue Lhomond\\ F-75231 Paris Cedex 05, France}
\author{Matthias Staudacher \footnote{matthias@aei-potsdam.mpg.de }
\footnote{Supported in part by EU Contract FMRX-CT96-0012} }
\address{
Albert-Einstein-Institut, Max-Planck-Institut f\"{u}r
Gravitationsphysik\\ Schlaatzweg 1\\  D-14473 Potsdam, Germany }
\maketitle
\begin{abstract}
We investigate one-matrix correlation functions for finite $SU(N)$
Yang-Mills integrals with and without supersymmetry. We propose
novel convergence conditions for these correlators which we determine
from  the one-loop perturbative effective action.  These conditions
are found to agree with non-perturbative Monte Carlo calculations for
various gauge groups and dimensions.  Our results yield important
insights into the eigenvalue distributions $\rho(\lambda)$ of these
random matrix models. For the bosonic models, we find that the
spectral densities $\rho(\lambda)$ possess moments of all orders
as $N \rightarrow \infty$. In the supersymmetric case, $\rho(\lambda)$
is a wide distribution with an $N-$independent asymptotic behavior
$\rho(\lambda) \sim \lambda^{-3},\lambda^{-7},\lambda^{-15}$ for
dimensions $D=4,6,10$, respectively.

\end{abstract}
\vspace{5.5cm}
\newpage
\narrowtext

Recently there has been renewed interest in dimensional reductions
of $SU(N)$ Yang-Mills theories.  It has been argued that
these ``toy models'', apart from some subtleties, allow to
recover the full unreduced theories in the large $N$ limit \cite{ek}.
The supersymmetric reductions are relevant to D-brane physics and
have also been used in various attempts to define non-perturbative
formulations of quantum gravity, supermembranes and superstrings.
For the  complete reduction, all space-time dependence is
eliminated from the gauge ``fields'', and the continuum Yang-Mills
path integral becomes an ordinary multi-matrix integral.  The
integrals appear to be ill-defined due to the flat directions
(i.e.~commuting matrices) in the action.  Recent calculations
have however uncovered that in many cases of interest these matrix
integrals converge and thus {\it do not need to be regulated}.
This was analytically found in the supersymmetric case for $SU(2)$
in \cite{Sestern}. In \cite{kns},\cite{ks} we found by Monte Carlo
methods that the absolute convergence persists for larger values
of $N$. Furthermore, we numerically established (and proved
analytically for $SU(2)$) the convergence properties of the bosonic
(non-supersymmetric) integrals. Perturbative one-loop estimates in
favor of  convergence were presented for the supersymmetric case
in \cite{ikkt2} and, recently, in \cite{nishi}, for the bosonic
case.  Mathematically rigorous proofs for $N>2$ are however still
missing.  For some applications of these integrals see
e.g.~\cite{Sestern},\cite{ikkt1}, \cite{kostov},\cite{hollo}.

The integrals in question are 
\begin{equation}
\CZ_{D,N}^{\CN}:=\int \prod_{A=1}^{N^2-1} 
\Bigg( \prod_{\mu=1}^{D} \frac{d X_{\mu}^{A}}{\sqrt{2 \pi}} \Bigg) 
\Bigg( \prod_{\alpha=1}^{\CN} d\Psi_{\alpha}^{A} \Bigg)
\exp \bigg[  \frac{1}{2} \Tr 
[X_\mu,X_\nu] [X_\mu,X_\nu] + 
\Tr \Psi_{\alpha} [ \Gamma_{\alpha \beta}^{\mu} X_{\mu},\Psi_{\beta}]
\bigg].
\label{susyint}
\end{equation}
They correspond to fully reduced $D$-dimensional Euclidean
$SU(N)$ Yang-Mills theory.
Here $\CN$ is the number of real supersymmetries, and for $\CN >0$
the only possible dimensions are $D=3,4,6,10$ corresponding to
$\CN=2,4,8,16$, respectively. If $\CN=0$ (no supersymmetry), {\it
a priori} all $D \geq 2$ are possible and we simply omit the terms
with Grassmann fields both from the measure and action of
eq.(\ref{susyint}).  For the detailed notation we refer to
\cite{kns},\cite{ks}.

In view of \cite{Sestern},\cite{kns},\cite{ks}
the necessary and sufficient conditions of existence for the 
integrals eq.(\ref{susyint}) then appear to be
\begin{eqnarray} 
\left.
\begin{array}{ccc}
D=4,6,10 & \quad {\rm and} \quad &  N \geq 2 
\end{array}
\right\} 
&\quad {\rm for} \;\;\; & \CN >0\cr 
 & & \cr
\left.
\begin{array}{ccc}
\quad D=3 & \qquad {\rm and} \quad & \;N \geq 4 \\
& & \\
\quad D=4 & \qquad {\rm and} \quad & \;N \geq 3 \\
& & \\
\quad D\geq5 & \qquad {\rm and} \quad &\;N \geq 2 
\end{array}  
\right\} 
&\quad {\rm for}\;\;\;& \CN=0
\label{intcond}
\end{eqnarray}
As already mentioned in \cite{ks}, it is only the bosonic $D=3$
$SU(3)$ integral which escapes a clear-cut classification by
numerical means as we cannot exclude almost marginal  convergence.
We have now checked that the susy $D=3$ integral is not absolutely
convergent\footnote{ This case, with $\CN=2$, is special in that,
at least for even $N$, the integral is formally zero\cite{kns}.
The integral is however not {\em absolutely} convergent, as we have
checked numerically with the methods used in this paper.  This
divergence is also present in the analytically tractable $SU(2)$
case. Adding a Chern-Simons term, as proposed in \cite{moore}, does
not improve convergence and the $D=3$ susy integral modified by
such a term is divergent for all $N$.} for any $N$, disproving
a conjecture made in \cite{ks}.

For the large $N$ limit, eq.(\ref{intcond}) implies existence of
the integrals for $D \geq 3$ in the bosonic case and for $D=4,6,10$
in the supersymmetric case.

Analytic motivation of eq.(\ref{intcond}) for general $N$ 
can be obtained from the one-loop perturbative calculations of
the effective action as presented in \cite{ikkt2},\cite{nishi}: 
Decompose the matrices into diagonal
$D(X_\mu)$ (with matrix elements $D(X_\mu)_{ij}=\delta_{ij} x_\mu^i$)
and off-diagonal components $Y_\mu$ (i.e.~$(Y_\mu)_{ii}=0$):
$X_\mu=D(X_\mu)+Y_\mu$. Fix a background gauge, 
expand the action to quadratic order in $Y_\mu$
and work out the effective action for the diagonal elements $x_\mu^i$.
The result for the supersymmetric case \cite{ikkt2} reads
\begin{equation}
\CZ_{D,N}^{\CN} \sim 
\int \prod_{i,\mu}^{N,D} dx_\mu^i~\Bigg[ \prod_\mu \delta\Big( 
\sum_i x_\mu^i \Big) \Bigg]~
~\sum_{G:\;{\rm maximal \; tree}}~
\prod_{(ij):\;{\rm link \; of }\; G}~{1 \over (x^i-x^j)^{3 (D-2)}} + \ldots
\label{susyfluct}
\end{equation}
Here the sum is over all possible maximal (i.e.~there are $N-1$ links in
$G$) trees $G$ connecting the $N$ coordinates.
As indicated in eq.(\ref{susyfluct}), there  are actually further terms 
present to 
one-loop order for $D=6,10$, but they are irrelevant for the powercounting
arguments below (see \cite{ikkt2}).
For the bosonic case \cite{nishi} one finds 
\begin{equation}
\CZ_{D,N}^{\CN=0} \sim \int \prod_{i,\mu}^{N,D} dx_\mu^i~\Bigg[ \prod_\mu 
\delta\Big( \sum_i x_\mu^i \Big) \Bigg]~ 
\prod_{i<j}~{1 \over (x^i-x^j)^{2 (D-2)}} 
\label{fluct}
\end{equation}
The one-loop  
approximation is reasonable for well-separated diagonal components
$x_\mu^i$ (that is the ``infrared'' regime). It is now easily verified
that superficial powercounting of all integrals on hyperplanes
of the integration space in eqs.(\ref{susyfluct}),(\ref{fluct})
results, at large $x_\mu^i$, precisely in the convergence conditions
eq.(\ref{intcond}). 

Clearly, this perturbative argument is {\it not} a proof
of eq.(\ref{intcond}): First, the cumulative effect of all corrections
to this one-loop effective action is beyond control. Second, there
are potentially dangerous regions in the integration space where
some of the differences $(x^i-x^j)^2$ are small (thus rendering
the one-loop approximation invalid) while others are large (leading
to a potential divergence).

An analytic calculation of the $D=4,6,10$ susy 
partition functions $\CZ_{N,D}^{\CN}$ has been reported in \cite{moore}.
In this work, light-cone variables $\phi=X_1+i X_D$, $\bar{\phi}=X_1-i X_D$
are used. Subsequently, $\phi$ and $\bar{\phi}$ are treated as independent
hermitian matrices (i.e.~one actually
has $\phi=X_1- X_0$, $\bar{\phi}=X_1+ X_0$), and the integrals computed in
\cite{moore} correspond to eq.(\ref{susyint}) with Minkowski (as opposed
to Euclidean) metric. However, the Minkowski integrals 
are divergent and need to be regulated. 
This calculation might be rendered rigorous
if one succeeded in deriving
the imposed pole prescriptions
directly from the matrix Wick rotation $X_D \rightarrow i X_0$.

For applications, it is important to understand
the statistical eigenvalue  distribution of these random matrix models.
The simplest quantity is the
distribution for the eigenvalues $\lambda_i$ 
of just one matrix, say, $X_1$, in
the background of the other matrices $X_2, \ldots, X_D$
\begin{equation}
\rho(\lambda)= {1 \over N} \bigg\langle \sum_{i=1}^N 
\delta(\lambda - \lambda_i) \bigg\rangle,
\label{density}
\end{equation}
Here, the average $<>$ is with respect to eq.(\ref{susyint}). The easiest
way to investigate the density is to consider the 
one-matrix correlators\footnote{
One might attempt to compute such correlators
within the powerful framework of \cite{moore}. The natural candidate
would be $\langle {1 \over N}\Tr \phi^k \rangle$ (where
$\phi=X_1 + i X_D$). Unfortunately, 
these correlators vanish for all $k$ in the Euclidean 
integral eq.(\ref{susyint}) 
due to the $SO(2)$ symmetry in the $(X_1,X_D)$ plane
(M.~Douglas, private communication).}
\begin{equation}
\Big\langle {1 \over N}\Tr X_1^{2 k} \Big\rangle = 
\int_{-\infty}^{\infty} d\lambda \rho(\lambda)~\lambda^{2 k}
\label{corr}
\end{equation}
In an ordinary hermitian matrix model, say with weight
$\exp(- \Tr X_1^2)$, the density eq.(\ref{density}) falls off
at infinity  as $\rho(\lambda) \sim \exp(- \lambda^2)$.
Therefore all moments eq.(\ref{corr}) exist. 

In the Yang-Mills ensembles for $SU(2)$, we obtained  analytically
(along the lines of \cite{Sestern}) that the correlators eq.(\ref{corr})
exist only for low values of $k$; we find $k<D-3$ ($\CN>0$) and
$k<{D \over 2}-2$ ($\CN=0$).  For general $N$, we obtain
the convergence conditions  in the one-loop approximations
eqs.(\ref{susyfluct}),(\ref{fluct}).  Some care has to be exercised,
since the most divergent contribution is not obtained by simply 
counting the overall
dimensionality of the integral. In fact, a more dangerous infrared
configuration stems from only a single (say, the $i$-th) eigenvalue
$x_\mu^i$ becoming large in the $D$-dimensional space. The only
exception being, once again, the bosonic $D=3$ $SU(3)$ integral.
Therefore, apart from this one case, we find
\begin{eqnarray}
\Big\langle {1 \over N}\Tr X_1^{2 k} \Big\rangle 
< \infty \;\;\;\;\;\; &{\rm iff}\;\;\;&
\left\{
\begin{array}{ccc}
k < D - 3 & \quad {\rm for} \quad &\; \quad \CN > 0 \\
&\\
k<N (D-2) - {3\over 2} D+2 & \quad {\rm for} 
\quad &  \quad \CN = 0
\end{array}
\right.
\label{corrcond}
\end{eqnarray}
This means that {\it no} moments exist for the $D=4$ susy integral,
and only the first two and six even moments for, respectively, 
the $D=6$ and $D=10$ susy integrals!   

We have been able to directly verify our conjecture eq.(\ref{corrcond}) 
with the 
Monte Carlo approach of \cite{kns},\cite{ks}. 
It may seem straightforward to obtain a direct numerical estimate
of the eigenvalue spectrum of $X_1$ by generating ensembles of 
$X= (X_1,X_2,\ldots,X_D)$  with the statistical weight
\begin{equation}
z_{D,N}^{\CN} (X) = \exp \bigg[  \frac{1}{2} \Tr
[X_\mu,X_\nu] [X_\mu,X_\nu] \bigg]~
\CP_{D,N}(X)
\label{integrand} 
\end{equation}
where $\CP_{D,N}(X)$ is the Pfaffian polynomial
coming from the integration over the fermionic degrees of freedom 
(cf \cite{kns}, for the bosonic integrals, this term is simply 
dropped).
The eigenvalue spectrum of $X$ can in principle be sampled and its
histogram generated. In practice, it is however impossible to
make firm predictions on the tails of the histogram, as they
comprise an exceedingly small number of samples.  For this reason,
we rather generate Markov chains of configurations $X(t)$  with
the statistical weight
\begin{equation}
\pi_{D,N}^{\CN,k}(X) = |\sum_\mu \Tr X_\mu ^{2 k} z_{D,N}^{\CN} (X)| 
\label{traceinteg}
\end{equation}
This means in particular that we perform one independent run for
every value of $k$.  The critical information on the convergence
condition is contained in the autocorrelation function of our
Markov-chain vector $X(t)$
\begin{equation}
f_t^{\Delta} = \frac{ X(t) \cdot X(t - \Delta)} 
{ 
\sqrt{ X(t  - \Delta) \cdot X(t - \Delta)} 
\sqrt{ X(t) \cdot X(t)} 
}    
\end{equation} 
for a fixed value of $\Delta$ (as in \cite{ks}, we have compactified
all the integrands). An absolutely non-integrable singularity of
the integrand corresponds to an infinite statistical weight
concentrated in a region of negligible extension. This simply means
that the simulation should get stuck, and the autocorrelation
function approach $1$.  An integrable singularity is not picked up
by the present method.
 
We have computed the autocorrelation function for a large number
of cases. Among the supersymmetric cases, we considered $N \le 6$
for $D=4$, $N \le 5$ for $D=6$ and $N = 2,3,4$ for $D=10$. All the
analytically known results for $SU(2)$ were easily reproduced.
Convergence of bosonic integrals was checked for $N \le 9 $ and $D
= 3,4,6$.

As an example, we show in Fig. 1 the  autocorrelation functions of
typical runs for the bosonic integrals with $D=3$, $N=6$  for $k=2$,
$k=3$ and $k=4$. The criterion eq.(\ref{corrcond}) leads us to
expect convergence for $k=2$ and $k=3$, as the smallest diverging
power is $k_{crit} = 3.5$. This behavior is clearly reproduced in
Fig. 1. Notice that the ``almost'' divergent case $k=3$ gets stuck
for very long periods of Monte-Carlo time, whereas the the almost
``convergent'' integral $k=4$ settles to a unit value of $f_t^{\delta}$
only after a very long transient.
\begin{center}
\begin{picture}(200,150)
\put(-150,0){ \includegraphics{632.eps} }
\put(0,0){ \includegraphics{633.eps} }
\put(150,0){ \includegraphics{634.eps} }
\end{picture}
\end{center}
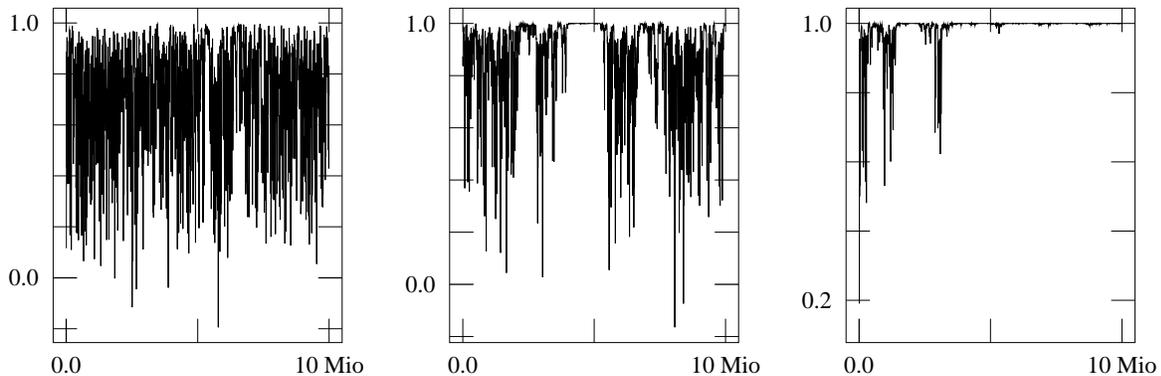
Fig. 1 {\it Autocorrelation functions $f_t^{\Delta}$ versus Monte Carlo
time $t$ ($\Delta=1000$) for the
bosonic integral with $N=6$, $D=3$, and, from the left, $k=2,3,4$.\\}

In Fig. 2, we show corresponding plots for the supersymmetric case
with $D=6$, $N=4$. Here, we expect the $k=2$ integral to be convergent,
as clearly found, while $k=4$ is divergent, as expected. In the marginally
divergent case, we have observed the typical alternations between 
``stuck'' and ``mixing'' behavior. 
\begin{center}
\begin{picture}(200,150)  
\put(-150,0){ \includegraphics{462.eps} }
\put(0,0){ \includegraphics{463.eps} }
\put(150,0){ \includegraphics{464.eps} }
\end{picture}
\end{center}
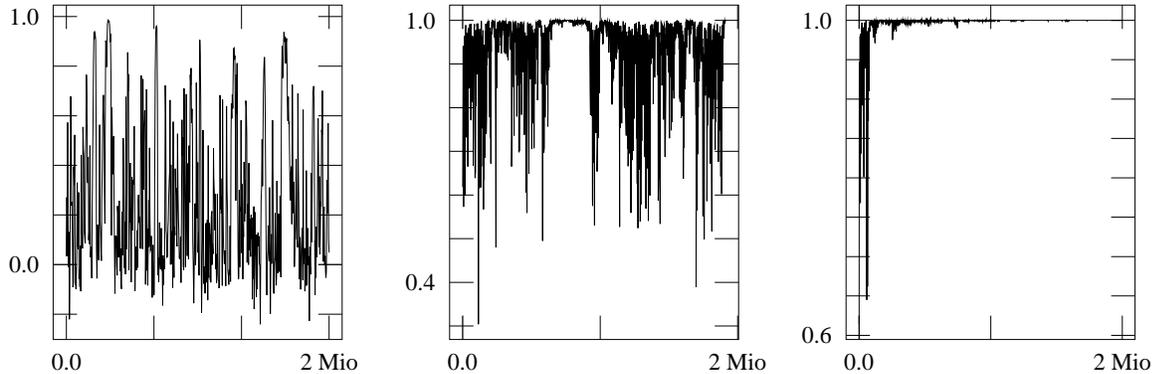
Fig. 2 {\it Autocorrelation functions $f_t^{\Delta}$ versus Monte Carlo
time $t$ ($\Delta=100$) for the 
supersymmetric integral with $D=6$, $N=4$, and, from the left, $k=2,3,4$.\\}

Notice that the middle plots of Figs 1 and 2 do not appear
fundamentally different, even though we expect the integrals to behave
differently. Our Monte Carlo procedure leaves a margin of error
which we estimate to be $\Delta k \sim \pm 1/2 $. A similar ambiguity
affects the abovementioned bosonic integral for $D=3$ and $N=3$.

The conditions eqs.(\ref{corrcond}) mean 
that in the Yang-Mills case the asymptotic behavior of the one-matrix 
eigenvalue distribution  
eq.(\ref{density}) decays algebraically:
\begin{equation}
\rho(\lambda) \sim |\lambda |^{-\alpha} \qquad {\rm for} \qquad
\lambda \rightarrow \pm \infty.
\label{asymp}
\end{equation}
Assuming this to be exactly true 
we can extract the power $\alpha$ in eq.(\ref{asymp}):
\begin{eqnarray}
\alpha =
\left\{
\begin{array}{ccc}
2 D-5 & \qquad {\rm for} & \CN>0 \\
&\\
2 N (D-2) -3 D +5 & \qquad {\rm for} & \CN=0 \\
\end{array}
\right.
\label{alpha}
\end{eqnarray}  
Most strikingly, the decay of the densities in the susy cases
$D=4,6,10$ ($\rho(\lambda) \sim \lambda^{-3},\lambda^{-7},\lambda^{-15}$)
is {\it independent} of $N$. It means that the eigenvalue distribution
are wide even in the  $N \rightarrow \infty$ limit!  This is a most 
unusual effect for a random matrix model.  Evidently, supersymmetry
is responsible for this behavior, as can e.g.~be seen by comparing
the one-loop effective actions for the diagonal matrix elements in
the susy ({\it cf} eq.(\ref{susyfluct})) and non-susy ({\it cf}
eq.(\ref{fluct})) case: In the latter, this effective action is
$\CO (N^2)$ while in the former it is $\CO(N)$.

The bosonic case is much more conventional in that the density
becomes concentrated in a finite interval at large $N$.  In this
context it is interesting to mention the result of a numerical
study of the bosonic Yang-Mills integrals which is complementary
to the present work:  In \cite{nishi}, Hotta et.al.~investigate
the scaling behavior of $\Big\langle {1 \over N}\Tr X_1^2 \Big\rangle
$ (i.e.~eq.(\ref{corr}) with $k=1$) with $N$. They find that, if
the variables in eq.(\ref{susyint}) are rescaled as $X_\mu^A
\rightarrow N^{{1 \over 4}} X_\mu^A$ such that an explicit factor
of $N$ appears in front of the action, this correlator tends to a
constant as $N \rightarrow \infty$. In consequence, the usual
't~Hooft scaling of large $N$ matrix models is obeyed, and 
the edge of the eigenvalue support tends to a constant. Nevertheless,
the observed asymptotic behavior ($\rho(\lambda) \sim \lambda^{- 2
N (D-2)}$ as $N \rightarrow \infty$) is different from the universal
exponential decay law of Wigner-type random systems.

To summarize, we have demonstrated how statistical information on
the eigenvalue distributions of Yang-Mills integrals (which are the
simplest examples for the so-called ``new'' matrix models) can be
obtained. We observed an interesting difference in the tails of
the spectral distribution between the susy and non-susy ensembles.
While suggestive -- e.g.~we find it interesting that the susy
eigenvalue distribution tends to stretch out much farther -- the
present study clearly does not yet address such important issues
as the full nonperturbative effective action for the diagonal
elements of all $D$ matrices or the problem of level spacing statistics.
These questions will need to be addressed if one intends
to apply Yang-Mills integrals to string theory and large $N$ gauge
theory.

\acknowledgements
We thank J.~Hoppe, V.~A.~Kazakov, I.~K.~Kostov,
H.~Nicolai and J.~Plefka for useful discussions.
M.~S. thanks the LPS-ENS Paris for hospitality. 
This work
was supported in part by the EU under Contract FMRX-CT96-0012.


\begin{references}

\bibitem{ek} T.~Eguchi and H.~Kawai,
{\it Reduction of dynamical degrees of freedom in the large N gauge theory},
Phys. Rev. Lett. 48 (1982) 1063.

\bibitem{Sestern} P.~Yi, {\it Witten Index and Threshold Bound States
of D-Branes}, Nucl.~Phys.~B505 (1997) 307, hep-th/9704098; 
S.~Sethi and M.~Stern, {\it D-Brane Bound State Redux}, hep-th/9705046.

\bibitem{kns} W.~Krauth, H.~Nicolai and M.~Staudacher,
{\it Monte Carlo Approach to M-Theory}, Phys.~Lett.~B431 (1998) 31, 
hep-th/9803117.

\bibitem{ks} W.~Krauth and M.~Staudacher,
{\it Finite Yang-Mills Integrals}, Phys.~Lett.~B435 (1998) 350, 
hep-th/9804199.

\bibitem{ikkt2} H.~Aoki, S.~Iso, H.~Kawai, Y.~Kitazawa, T.~Tada,
{\it Space-Time Structures from} IIB {\it Matrix Model}, 
Prog. Theor. Phys. 99 (1998) 713, hep-th/9802085.

\bibitem{nishi} T.~Hotta, J.~Nishimura, A.~Tsuchiya,
{\it Dynamical Aspects of Large $N$ Reduced Models}, hep-th/9811220.

\bibitem{ikkt1} N.~Ishibashi, H.~Kawai, Y.~Kitazawa, A.~Tsuchiya,
{\it A Large} $N$ {\it Reduced Model as Superstring},
Nucl. Phys. B498 (1997) 467, hep-th/9612115.

\bibitem{kostov} I.K.~Kostov and P.~Vanhove, {\it Matrix String Partition
Functions}, Phys. Lett. B444 (1998) 196, hep-th/9809130.

\bibitem{hollo} N.~Dorey, T.J.~Hollowood, V.V.~Khoze, M.P.~Mattis, 
S.~Vandoren, {\it Multi-Instantons and Maldacena's Conjecture}, hep-th/9810243;
{\it Multi-Instanton Calculus and the AdS/CFT Correspondence in} $\CN=4$
{\it Superconformal Field Theory}, hep-th/9901128.

\bibitem{moore} G.~Moore, N.~Nekrasov and S.~Shatashvili,
{\it D-particle bound states and generalized instantons}, hep-th/9803265.



\end{references}
\end{document}